\documentclass[aps,prl,twocolumn,superscriptaddress,showpacs]{revtex4-1}

\usepackage{graphicx}
\usepackage{CJK}
\usepackage{setspace}

\begin{document}


\title{Quantum and Thermal Transitions Out of the Pair-Supersolid Phase of Two-Species Bosons in Lattice}



\author{Chia-Min Chung}
\author{Shiang Fang}
\affiliation{Department of Physics, National Tsing Hua University, Hsinchu 30013, Taiwan}

\author{Pochung Chen}
\email[]{pcchen@phys.nthu.edu.tw}
\affiliation{Department of Physics, National Tsing Hua University, Hsinchu 30013, Taiwan}
\affiliation{Frontier Research Center on Fundamental and Applied Sciences of Matters, 
National Tsing Hua University, Hsinchu 30013, Taiwan}


\date{\today}

\begin{abstract}
We investigate two-species bosons in a two-dimensional square lattice by quantum Monte Carlo method. 
We show that the inter-species attraction and nearest-neighbor intra-species repulsion 
results in the pair-supersolid phase, where a diagonal solid order coexists with an off-diagonal
pair-superfluid order. The quantum and thermal transitions out of the pair-supersolid phase are 
characterized. It is found that there is a direct first order transition from the pair-supersolid phase
to the double-superfluid phase without an intermediate region. Furthermore, the melting
of the pair-supersolid occurs in two steps. Upon heating, first the pair-superfluid is destroyed 
via a Kosterlitz-Thouless transition then the solid order melts via an Ising transition.

\end{abstract}


\maketitle


Supersolid (SS) is an intriguing but counter-intuitive concept, for it is characterized 
by the co-existence of a diagonal long-range order (DLRO) of solid and an off-diagonal 
long-range order (ODLRO) of superfluid (SF). The possible existence of the SS phase 
has been studied experimentally and theoretical since the early 1970's 
\cite{Leggett1970,Chester1970,Kim2004}. 
In recent years, numerical simulations
have established the existence of the SS phase in a number of interacting
lattice boson models \cite{Sengupta2005}. Recently, the concept of pair-supersolid (PSS) for
two-species lattice bosons is proposed \cite{Trefzger2009}.
Here the pair-superfluid (PSF) replaces the SF to be the off-diagonal long-range order.
In the PSF phase each species individually doesn't form the SF, 
instead the boson-boson composites form the PSF. It has been shown theoretically that
PSF phase arises naturally in bosonic mixtures with inter-species 
attractions \cite{Altman2003,Kuklov2004a,Chen2010f}.
Some of the optimal candidates to realize the PSF phase include
bosonic binary mixtures in optical lattices and dipolar particles in bilayer optical lattices.
The later system is also an ideal candidate to realize the PSS phase because the dipole-dipole
interaction can provide both an interlayer attraction and an intralayer nearest neighbor repulsion
which give rise to the PSS phase \cite{Trefzger2009}.

In this work we consider the two-species Bose-Hubbard model in a two-dimensional (2D)
square lattice, which is characterized by the Hamiltonian:
\begin{eqnarray}
  \label{eqn:H}
  H&=& \nonumber
  -t\sum_{\langle ij\rangle} \left( a^{\dagger}_i a_j+b^{\dagger}_i b_j +h.c. \right)
  +V \sum_{\langle ij\rangle \sigma}  n^\sigma_i n^\sigma_j \\
  &+& \frac{U}{2} \sum_{i\sigma} n^{\sigma}_i \left( n^{\sigma}_i-1 \right) +W \sum_i n^a_i n^b_i 
  -\mu \sum_{i\sigma} n^\sigma_i,
\end{eqnarray}
where $\sigma=a,b$ indicates the two species respectively,
$U$ ($W$) is the on-site intra-species (inter-species) interaction, and
$V>0$ is the intra-species nearest neighbor repulsion. 
It is known that without the intra-species repulsion $V$, the Hamiltonian gives rise to 
PSF phase for $W<0$ \cite{Kuklov2003,Kuklov2004a,Arguelles2007} 
and super-counterfluid (SCF) phase for $W>0$ \cite{Kuklov2003}, in addition to 
the conventional Mott insulator and double-superfluid (2SF) phases. It has been proposed that
the inclusion of intra-species repulsion $V$ results in the checkerboard solid and
PSS phases \cite{Trefzger2009}. 
In Ref.~\cite{Trefzger2009} the existence of such a PSS phase is predicted by
solving the time dependent Gutzwiller equations for the low-energy effective Hamiltonian of boson pairs.
However, the stability analysis of the PSS phase against phase separation is beyond such a treatment.
Furthermore, the nature of the associated quantum phase transitions remains unknown and the precise
phase boundaries are not determined.
It is thus imperative to study the phase diagram beyond Gutzwiller approximation
and to investigate the nature of associated quantum phase transitions. 
Additionally, so far the theoretical investigations of the PSS phase have focused on 
it's existence and the stability at zero temperature.
Since it is known that typically the thermal melting of a SS state contains two distinct 
transitions \cite{Boninsegni2005,Laflorencie2007},
a precise characterization of the thermal transition out of the PSS state is also called for.

\begin{figure}[btp]
\begin{center}
\includegraphics[width=0.9\columnwidth]{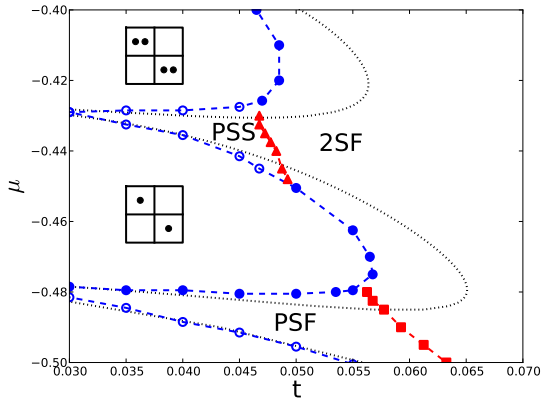}
\caption{(Color online)
Ground state phase diagram for the two-species Bose-Hubbard model for $W=-0.95U$ and $V=0.025U$. 
First-order (continuous) transitions are shown by filled (open) symbols
and are estimated using data with $L=30$ ($L=24$).
The black dotted lines indicate the phase boundaries determined by the mean-field solution
as described in Ref.~\cite{Trefzger2009}.
The density profile of the checkerboard solid lobes are sketched inside the lobe.
}
\label{fig:phases}
\end{center}
\end{figure}

\begin{figure}[tbp]
\begin{center}
\includegraphics[width=0.9\columnwidth]{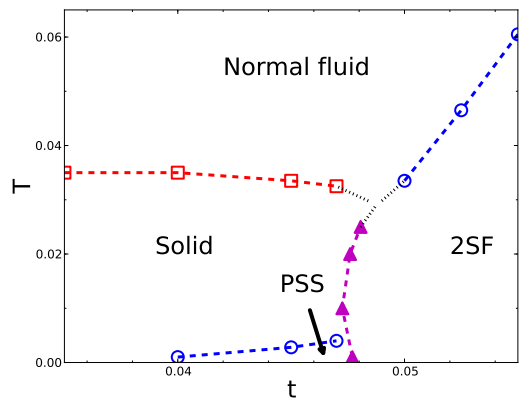}
\caption{(Color online)
Finite temperature phase diagram for the two-species Bose-Hubbard model 
for $W=-0.95U$, $V=0.025U$, and $\mu=-0.435U$. 
First-order (Ising) transition is shown by filled triangle (open square),
while KT transitions are shown by open circles.
Finite-size scaling is used to determine the critical temperature as described in the text.
Black dotted lines represent the estimated phase boundaries near the multicritical point.
Lines are guides to the eyes.
}
\label{fig:Tphases}
\end{center}
\end{figure}

In this work we perform large-scale quantum Monte-Carlo (QMC) simulations to study the quantum
and thermal phase transitions of the two-species Bose-Hubbard model characterized by Eq.~(\ref{eqn:H}).
In particular, we employ a multi-worm algorithm which is similar to the method proposed in Ref.~\cite{Ohgoe2011}.
We shall briefly summarize our main results before we discuss them in detail: 
We obtain the ground state phase diagram in the $\mu, t$ plane as shown in Fig.~\ref{fig:phases}.
{
Two checkerboard solid lobes with density $n_\sigma=1/2$ and $n_\sigma=1$ respectively are identified.
Here $n_\sigma=\sum_i \langle n^\sigma_i\rangle/L^2$ is the average particle density of $\sigma$-species 
boson and $L$ is the system size.
}
We find that the PSS phase appears between these two lobes while the PSF phase appears between 
the $n_\sigma=1/2$ lobe and vacuum. We observe that all transitions into the 2SF phase are first-order in the 
parameter region we studied. These include the PSS-2SF, PSF-2SF, solid-2SF, and vacuum-2SF transitions.
In contrast, the PSS-solid and PSF-vacuum transitions are continuous. But the PSF to $n_\sigma=1/2$ 
solid transition is first-order. Furthermore, we establish that the transition from PSS to 2SF is 
direct without an intermediate region. 
We also obtain the finite temperature phase diagram on the $t, T$ plane for 
a fixed $\mu=-0.435$ as shown in Fig.~\ref{fig:Tphases}.
We find that the melting of PSS phase contains two distinct transitions: 
Upon heating first the PSF is destroyed via a KT transition. 
When heated further the solid order melts via an Ising transition.
Finally, the finite-temperature solid to 2SF transition is first order. 

In the following we discuss our simulations and findings in more detail.
In all calculations we set $U=1$ as our energy unit. 
We note that the system can exhibit three kinds of ODLRO corresponding 
to three kinds of superfluid density.
It is known that single species SF density $\rho^\sigma_s$ can be estimated by 
$\rho^\sigma_s=\langle \left(\mathbf{W}_\sigma\right)^2 \rangle/(4\beta  t)$,
where $\mathbf{W}_\sigma$ is the winding number for $\sigma$ species boson and 
$\beta$ is the inverse temperature \cite{Pollock1987}.
The PSF (SCF) density can be identified 
via the sum (difference) of the winding number with the formula
$\rho^{\text{PSF(SCF)}}_s = \langle \left(\mathbf{W}_\pm \right)^2 \rangle/(2\beta t)$,
where $\mathbf{W}_\pm=(\mathbf{W}_a \pm \mathbf{W}_b)/2$.
The PSF phase is identified by a finite PSF density
($\rho^{\text{PSF}}_s \neq 0$) together with a zero SCF density ($\rho^{\text{SCF}}_s =0$),
which characterizes a perfectly correlation between bosons of different species. 
The system is in a 2SF phase if all three kinds of superfluid densities are nonzero.
For small enough $t$, all kinds of superfluid densities are zero and the ground state may acquire a DLRO, 
which can be detected by measuring the structure factor 
\begin{equation}
  S_\sigma \equiv \frac{1}{N} \sum_{ij} e^{\left(\pi,\pi\right) \cdot \left(\mathbf{r}_i-\mathbf{r}_j\right)}\langle n^\sigma_i n^\sigma_j\rangle.
\end{equation}
Due to the symmetry of the model one has $S_a\approx S_b$ and the index $\sigma$ can be dropped for simplicity.
The system is in the PSS phase if the structure factor is nonzero
while the system also has perfectly correlated superfluidity 
($\rho^{\text{PSF}}_s \neq 0$ and $\rho^{\text{SCF}}_s =0$).

When $U+W,V\ll U$, a low-energy effective Hamiltonian of pairs can be derived using second order perturbation theory. 
Within the mean-field theory the paired order parameter can be determined by solving the 
self-consistency equation \cite{Trefzger2009}.
When $U+W<4V$ the mean-filed analysis predicts that both the $n_\sigma=1/2$ and $n_\sigma=1$ 
insulating lobes have checkerboard order and PSS/PSF phases may exist outside the lobes.
To firmly establish the existence of the PSS phase and to compare precisely with the results in Ref.~\cite{Trefzger2009}
we set $W=-0.95$ and $V=0.025$ throughout this work.
As shown in Fig.~\ref{fig:phases}, we find that mean-field results substantially deviate from our QMC results
and mean-filed solutions overestimate the region of solid phase by a large margin.
The PSS/PSF regions identified by QMC are smaller than the results by time-dependent Gutzwiller equations 
and no PSS region is found below the $n_\sigma=1/2$ lobe.
In addition for the first lobe we observe a very weak
reentrant behavior, which is similar to what has been found in 1D,
but such a reentrant behavior is absent for higher lobes \cite{Arguelles2007}.

\begin{figure}[tbp]
\begin{center}
\includegraphics[width=0.9\columnwidth]{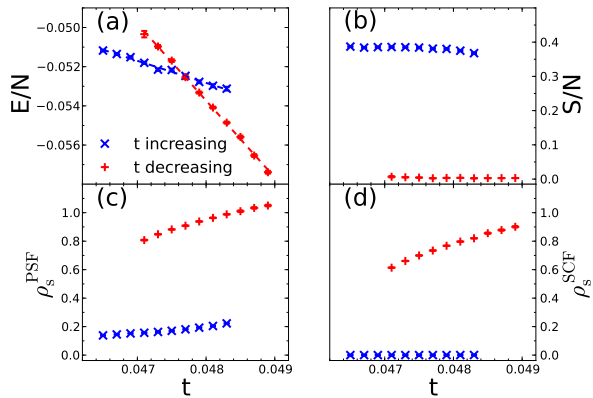}
\caption{(Color online) 
(a) Energy per site $E/N$, (b) structure factor $S/N$, (c) PSF density $\rho^{\text{PSF}}_s$, 
and (d) SCF density $\rho^{\text{SCF}}_s$ for $\mu=-0.435U$, $\beta=1000$, and $L=24$.
The hysteresis effects were obtained by increasing (decreasing) $t$ 
in small steps starting from a configuration inside the PSS (2SF) phase.
The last configuration generated in the run at a particular $t$ are used as the starting configuration
for the next value of $t$.
The critical hopping $t_c$ is extracted from the crossing of the energy branches.
The QMC error bars are smaller than the symbols.
}
\label{fig:mu0.44}
\end{center}
\end{figure}

Now focusing on the quantum transitions out of the PSS phase.
Similar to the continuous PSF-vacuum transition, we find that the PSS-solid transition is also continuous. 
The PSS-2SF transition, however, is strongly first order.
Near the first order phase boundary the Markovian dynamics of the simulation suffers from
the critical slowing down due to the large tunneling time between the competing phases.
Interestingly, it is pointed out recently that such a large tunneling time can be used to locate the phase boundary 
accurately by determining the average energy of the two different phases in the proper manner \cite{Sen2010a}.
We first determine the energy of the PSS phase by starting the simulation at a value of $t$ inside 
the PSS phase and then increase $t$ in small steps to go to the 2SF phase.
During this procedure we always use the last configuration generated in the run
at a particular $t$ as the starting configuration for the next value of $t$.
For large enough system the tunneling time is longer than the simulation time and 
the procedure ensures that the system stays in the metastable PSS state even when the 
phase boundary has been crossed.
By starting from a value of $t$ inside the 2SF phase and decrease $t$ in the same fashion, 
the energy of the 2SF phase is obtained. The crossing point of the two energy branches then gives $t_c$.
In Fig.~\ref{fig:mu0.44}(a) we show the energy per site for $\mu=-0.435$, $\beta=1000$, and $L=24$.
The crossing point of two energy branches gives an estimation of $t_c\simeq 0.0477(2)$.
In Fig.~\ref{fig:mu0.44}(b), (c), (d) we also show the measurements of $S/N$, $\rho^{\text{PSF}}_s$, 
and $\rho^{\text{SCF}}_s$ respectively, where the hysteresis behavior is clearly observed.


\begin{figure}[tbp]
\begin{center}
\includegraphics[width=0.9\columnwidth]{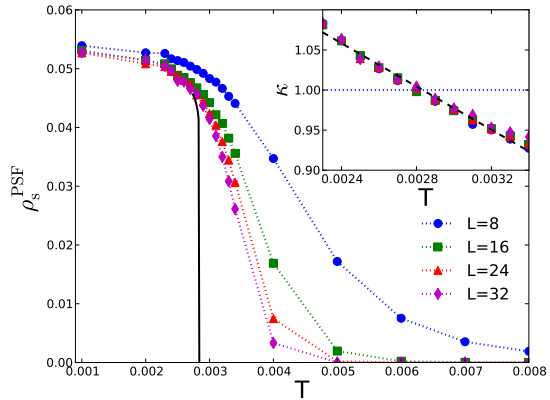}
\caption{(Color online)
PSF density in the vicinity of the KT transition temperature $T^{\text{PSF}}_{\text{KT}}$ 
for $t=0.045$ with system sizes $8 \leq L \leq 32$ and the infinite system (solid line).
Inset: $\kappa(T)$ in Eq.~(\ref{eq:RG}) for different pairs of system sizes $(L_1,L_2)=(8,16)$(blue circle), 
$(8,24)$(green squre), $(8,32)$(red triangle) and $(16,32)$(magenta diamond).
The dash line shows the liner fitting for 
$\kappa(T)=1+c(T^{\text{PSF}}_{\text{KT}}-T)$ with $T^{\text{PSF}}_{\text{KT}}\simeq 0.00283(7)$ and 
$c\simeq 134.88(1)$ respectively.
The QMC error bars are smaller than the symbols.
}
\label{fig:PSF-NF}
\end{center}
\end{figure}

We now turn to the finite temperature properties. 
For PSS phase, it is expected that the solid order can persist up to some finite temperature $T_c$, 
beyond which the solid order melts due to thermal fluctuation. 
On the other hand, while in 2D one cannot 
break the continuous symmetry at finite temperature, quasi-long-range order can persist up
to a Kosterlitz-Thouless (KT) transition temperatures. 
Consequently the PSF density $\rho^{\text{PSF}}_s$ remains finite up to the 
critical temperature $T^{\text{PSF}}_{\text{KT}}$ and 
a persistence of the PSS state at finite temperature is expected. However, it is not 
straightforward to see whether the solid order and the PSF density disappear simultaneously, 
or which order will be destroyed earlier by thermal fluctuation.
It is also not clear whether an intermediate SS or PSF region can appear at finite temperature.
Furthermore, it is expected that for 2SF phase the superfluid density $\rho^\sigma_s$ should also remain 
finite up to the critical temperature $T^\sigma_{\text{KT}}$, 
but it is not clear whether $T^\sigma_{\text{KT}}$ will connect smoothly to
$T^{\text{PSF}}_{\text{KT}}$ as one varies the system parameters across the PSS-2SF phase boundary.

To accurately determine the KT transition temperature we utilize the KT renormalization group equations. 
It is known that for an infinite system there should be a universal jump 
at the transition temperature. For finite-size system, however, the transition is smeared by logarithmic finite-size effects.
Here we follow the proposed method in Ref.~\cite{Boninsegni2005} to determine 
$T^\sigma_{\text{KT}}$ $(T^{\text{PSF}}_{\text{KT}})$ in the thermodynamic limit. 
We define $R^\sigma_s(L,T) \equiv t \pi \rho^\sigma_s(L)/T$ $(R^{\text{PSF}}_s(L,T) \equiv t \pi \rho^{\text{PSF}}_s(L)/2T)$
and study the finite-size
scaling using KT renormalization group equations in the integral form \cite{Prokofev2002}
\begin{equation}
  4\ln(L_2/L_1)=\int_{R(L_2,T)}^{R(L_1,T)}\frac{dt}{t^2 (\ln(t)-\kappa(T)) +t}.
  \label{eq:RG}
\end{equation}
Here $\kappa(T)$ is a system size independent parameter, which is analytic in terms of temperature.
It is expected that $\kappa(T)\simeq 1+c(T_{\text{KT}}-T)$ for $T <T_{\text{KT}}$.
For each pair of system sizes, one can determine a curve for $\kappa(T)$.
The data supports a transition of the KT type if $\kappa(T)$ obtained
from different pairs of the system sizes collapse into a straight line around $\kappa=1$.
The transition temperature is then determined by $\kappa(T_{\text{KT}})=1$
and the superfluid density in thermodynamic limit can be determined by the equation $1/R+\ln R=\kappa(T)$.

\begin{figure}[tbp]
\begin{center}
\includegraphics[width=0.9\columnwidth]{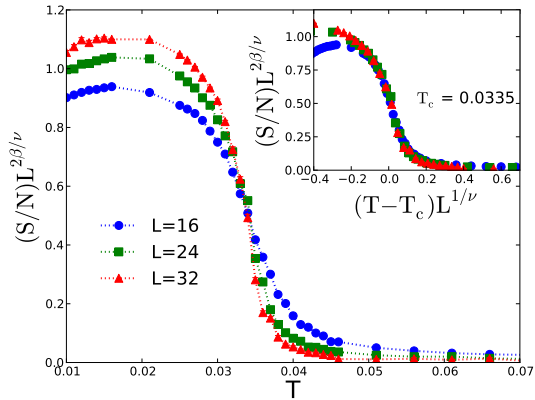}
\caption{(Color online) 
Rescaled structure factor $S(\pi,\pi)L^{2\beta/\nu}/N$ versus temperature for various system sizes
near the solid melting temperature. 
The crossing point indicates the critical temperature $T_c\simeq0.0335(5)$. 
Inset: Data collapse with rescaled temperature $(T-T_c)L^{1/\nu}$.
The Ising critical exponents $\beta=1/8$ and $\nu=1$ are used.
The QMC error bars are smaller than the symbols.
}
\label{fig:S-NF}
\end{center}
\end{figure}

In Fig.~\ref{fig:PSF-NF} we show the PSF density as a function of temperature for different 
system sizes for $t=0.045$ and $\mu=-0.435$. In the inset we show $\kappa(T)$ obtained from different pairs of 
system sizes. The data collapse and the smooth analytic behavior of $\kappa(T)$ confirm the KT nature of the
transition, and the transition temperature is found to be $T^{\text{PSF}}_{\text{KT}}\simeq 0.00283(7)$. 
The PSF density in thermodynamic limit with universal jump is plotted as solid line. 
Similar analysis were applied to the 2SF-normal fluid transition. 
By identifying $T^\sigma_{\text{KT}}$ and $T^{\text{PSF}}_{\text{KT}}$ at different $t$
we find that in general $T^\sigma_{\text{KT}}$ is much higher than the $T^{\text{PSF}}_{\text{KT}}$, 
as shown in Fig.~\ref{fig:Tphases}.
They don't smoothly connect to each other as $t$ is varied across the zero temperature PSS-2SF phase boundary.

In Fig.~\ref{fig:Tphases} we plot the finite temperature phase diagram on $t-T$ plane
for the cut at $\mu=-0.435$. We find that the melting
of PSS occurs in two distinct steps: the PSF is destroyed before 
the melting of the solid order occurs. 
We note that PSF and SF density disappear simultaneously 
while the solid order is still very strong above $T^{\text{PSF}}_{\text{KT}}$. 
This indicates that there is a direct PSS-Solid transition without an intermediate SS phase.
We also observe that the finite-temperature solid-2SF transition is a direct first order transition
without an intermediate region. The first order nature is expected due to the different 
symmetries in different phases. 

Finally we study the melting of the sold phase. It is expected that at the critical temperature $T_c$ the solid
melts via 2D Ising transition. To confirm the Ising nature of the transition we use the 
finite size scaling hypothesis for the structure factor $S/N=L^{-2\beta/\nu}\mathcal{F}((T-T_c) L^{1/\nu})$.
The Ising universality class can be checked via the data collapse for $(S/N) L^{2\beta/\nu}$ v.s. 
$(T-T_c) L^{1/\nu}$ for various $L$ with the Ising exponent $\beta=1/8$ and $\nu=1$.
In Fig.~\ref{fig:S-NF} we plot the re-scaled structure factor as as a function of the temperature,
where the crossing point provides an estimation of the critical temperature $T_c$.
Data collapse is clearly observed if the temperature is also rescaled, as shown in the inset 
of Fig.~\ref{fig:S-NF}. The Ising nature of the transition is hence clearly confirmed.

\section{Summary}
In summary we have studied the two-species Bose-Hubbard model with on-site intra-species (inter-species) 
repulsion (attraction) and nearest neighbor repulsion in a 2D square lattice.
We obtain the ground state and finite temperature phase diagrams and 
firmly establish the existence of PSF and PSS phases within this model. 
We accurately determine the phase boundaries and the nature of the phase transitions.
Interestingly, we find that the transition from PSS to 2SF or solid phase is direct without an intermediate phase.
Furthermore, the melting of the PSS occurs in two steps:
the PSF is first destroyed via a KT transition then the solid order melts via an Ising transition.


\begin{acknowledgments}
We acknowledge the support from NCHC, NCTS and NSC Taiwan.
We thank Min-Fong Yang, Fu-Jiun Jiang, Barbara Capogrosso-Sansone, Ying-Jer Kao, and Naoki Kawashima 
for useful discussions.
\end{acknowledgments}

\bibliography{PSS}

\end{document}